\documentclass[12pt]{article}
\usepackage[total={6in, 9in}]{geometry}
\usepackage{amssymb,amsthm}
\usepackage{natbib}
\usepackage{graphicx}
\usepackage{fancyref}
\usepackage{multirow}
\usepackage{amsmath,amsfonts}
\usepackage{longtable}
\usepackage{color}
\usepackage{bm}
\usepackage{lmodern}
\usepackage{epstopdf}
\usepackage{indentfirst}

\usepackage{subfigure}
\usepackage{overpic}
\usepackage{supertabular}
\usepackage{array}
\usepackage{url}
\usepackage{float}    
\usepackage{setspace}
\usepackage{enumerate}  

\usepackage{rotating} 
\usepackage{lscape}  
\usepackage{extarrows}    

\newenvironment{sequation*}{\begin{equation*}\footnotesize}{\end{equation*}}
\usepackage[T1]{fontenc}
\usepackage[utf8]{inputenc}
\usepackage{authblk}
\usepackage{tikz}
\usetikzlibrary{shapes,arrows,chains}	
\usetikzlibrary{decorations.pathmorphing, backgrounds, positioning, fit, petri, automata}
\usepackage{xcolor}     
\definecolor{yellow1}{rgb}{1,0.8,0.2}      
\definecolor{LightBlue1}{RGB}{202,225,255}         
\definecolor{SteelBlue3}{RGB}{79,148,205}

{\bf}{\it}

{\bf}{\it}

\newcommand{\bfA}{\mbox{\boldmath $A$}}
\newcommand{\bfB}{\mbox{\boldmath $B$}}

\newcommand{\bfw}{\mbox{\boldmath $w$}}

\newcommand{\FS}[2]{\displaystyle\frac{#1}{#2}}

\title{Fixed-effects model: the most convincing model for meta-analysis with few studies}
\author[1]{Enxuan Lin}
\author[1,\footnote{Corresponding author. E-mail: tongt@hkbu.edu.hk}]{Tiejun Tong}
\author[2]{Yong Chen}
\author[3]{Yuedong Wang}
\affil[1]{\small Department of Mathmatics, Hong Kong Baptist University, Hong Kong}
\affil[2]{Department of Biostatistics and Epidemiology, University of Pennsylvania, Philadelphia, USA}
\affil[3]{Department of Statistics and Applied Probability, University of California, Santa Barbara, USA}
\date{}
\begin{document}
\maketitle
\begin{abstract}
\baselineskip 17pt
\noindent
According to \cite{davey2011characteristics} with a total of 22,453 meta-analyses from the January 2008 Issue of the Cochrane Database of Systematic Reviews, the median number of studies included in each meta-analysis is only three.
In other words, about a half or more of meta-analyses conducted in the literature include only two or three studies.
While the common-effect model (also referred to as the fixed-effect model) may lead to misleading results when the heterogeneity among studies is large, the conclusions based on the random-effects model may also be unreliable when the number of studies is small.
Alternatively, the fixed-effects model avoids the restrictive assumption in the common-effect model and the need to estimate the between-study variance in the random-effects model.
We note, however, that the fixed-effects model is under appreciated and rarely used in practice until recently.
In this paper, we compare all three models and demonstrate the usefulness of the fixed-effects model when the number of studies is small.
In addition, we propose a new estimator for the \textit{unweighted average effect} in the fixed-effects model.
Simulations and real examples are also used to illustrate the benefits of the fixed-effects model and the new estimator.

\end{abstract}	
\baselineskip 21pt
\newpage

\section{Introduction}
\setlength{\baselineskip}{22pt}
\vspace{-1em}
\noindent
Meta-analysis is a quantitative method for synthesizing results from multiple independent studies.
Through combining the randomized trials of the same comparison of interventions, the estimates of correlation between two variables, or the measures of the accuracy of a diagnostic test, meta-analysis can help researchers to understand inter-study differences, to compare multiple treatments simultaneously,
and, more generally, to synthesize evidence from different types of studies (\citealp{berkey1995random}; \citealp{hunter2004methods};  \citealp{lu2004combination}; \citealp{ades2006multiparameter};  \citealp{leeflang2008systematic}; \citealp{borenstein2011introduction}).
According to Web of Science Core Collection as of 12 December 2019, there are more than 186,600 publications in the literature with the keyword of ``meta-analyses'' in the title or abstract.
Without any doubt, meta-analysis has become one of the most important areas in medical research.

There are two commonly used  models in meta-analysis: the common-effect model and the random-effects model.
The common-effect model, also known as the fixed-effect model, assumes one common effect across all the studies, denoted by the \textit{common effect}.
In contrast, the random-effects model assumes that the true effect sizes may vary from study to study but they follow a certain distribution, e.g. a normal distribution.
The summary effect in the random-effects model is known as the \textit{mean effect}, which is the mean of the underlying distribution.

According to \cite{davey2011characteristics} with a total of 22,453 meta-analyses from the January 2008 Issue of the Cochrane Database of Systematic Reviews, the median number of studies included in each meta-analysis is only three.
In other words, about a half or more of  meta-analyses conducted in the literature include only two or three studies.
For meta-analysis with few studies in the presence of heterogeneity, the common-effect model and the random-effects model are likely to yield misleading or unreliable results (\citealp{hedges1998fixed}; \citealp{poole1999random}; \citealp{henmi2010confidence}; \citealp{gonnermann2015no}; \citealp{rover2015hartung}; \citealp{bender2018methods}).
On one hand, when the observed effect sizes have large differences, the common-effect model often has an inflated  type \uppercase\expandafter{\romannumeral1} error rate and provides a poor coverage probability  for estimated confidence interval of the \textit{common effect}.
On the other hand, the estimate of the \textit{mean effect}  in the random-effects model relies heavily on the between-study variance estimate.
For meta-analysis with few studies, however, it is unlikely to have a reliable estimate for the between-study variance(\citealp{cornell2014random}).

In addition to the two common models, there is another model available for meta-analysis termed as the fixed-effects model (\citealp{laird1990some}; \citealp{hedges1998fixed}).
The fixed-effects model assumes that the effect sizes of the studies are deterministic and different.
That is, it neither assumes the study-specific effect sizes to be all the same as in the common-effect model, nor requires them to follow a certain distribution as in the random-effects model.
For meta-analysis with few studies, the fixed-effects model is capable to provide a good balance between the common-effect model and the random-effects model.
It not only avoids the unreasonable assumption
on a \textit{common effect} in the common-effect model when the heterogeneity exists, but also avoids the low accuracy of the between-study variance estimate in the random-effects model when there are only few studies.

For the fixed-effects model, two summary effects have been suggested in the literature: the \textit{unweighted average effect} (\citealp{laird1990some}; \citealp{bender2018methods})  and the  \textit{weighted average effect} (\citealp{rice2018re}).
The summary effects selection should depend on the research goal of the meta-analysis.
If  all the studies are regraded as equally important, the \textit{unweighted average effect} by \cite{laird1990some} should be used.
We note that the summary effects and conclusions for the fixed-effects models are pertaining to the observed studies only.
The estimator of the \textit{weighted average effect} proposed by \cite{rice2018re} is the same as the estimator of the \textit{common effect} in the common-effect model.
For the \textit{unweighted average effect}, the existing unbiased estimator by \cite{laird1990some} does not fully utilize the information across the studies.

In this paper, we demonstrate that the fixed-effects model can serve as the most convincing model for meta-analysis with few studies through real examples.
For the \textit{unweighted average effect} in the fixed-effects model, in addition to the unweighted average, we propose a new estimator with different weight for each study that fully utilizes the information across the studies.
Through theory and simulation, we demonstrate that our optimal estimator performs better than the unweighted average estimator in all settings in the sense of having minimal mean squared error.

The rest of the paper is organized as follows.
In Section 2, we review the three models for meta-analysis.
In Section 3, we propose an optimal estimator for the \textit{unweighted average effect} and assess its performance through numerical studies.
In Section 4, we illustrate the usefulness of the fixed-effects model using real examples.
Finally, we conclude the paper with a brief discussion in Section 5, and provide the technical results in the Appendix.

\vskip 12pt
\section{Existing models for meta-analysis}
\vspace{-1em}
\noindent
The common-effect model and the random-effects model are two most popular models for meta-analysis.
They were first introduced  in the 1950s (\citealp{bliss1952statistics}; \citealp{cochran1954combination}), and have now been widely adopted for synthesizing multiple studies.
The fixed-effects model was introduced in \cite{laird1990some} and \cite{hedges1998fixed}, but received little attention until recently (\citealp{bender2018methods}; \citealp{rice2018re}).
In this section, we provide a brief review on the three models, including their model assumptions, their effect sizes, and their key limitations in performing a meta-analysis.

\begin{figure}
	\centering
	\includegraphics[width=	1\linewidth, height=0.6\textheight]{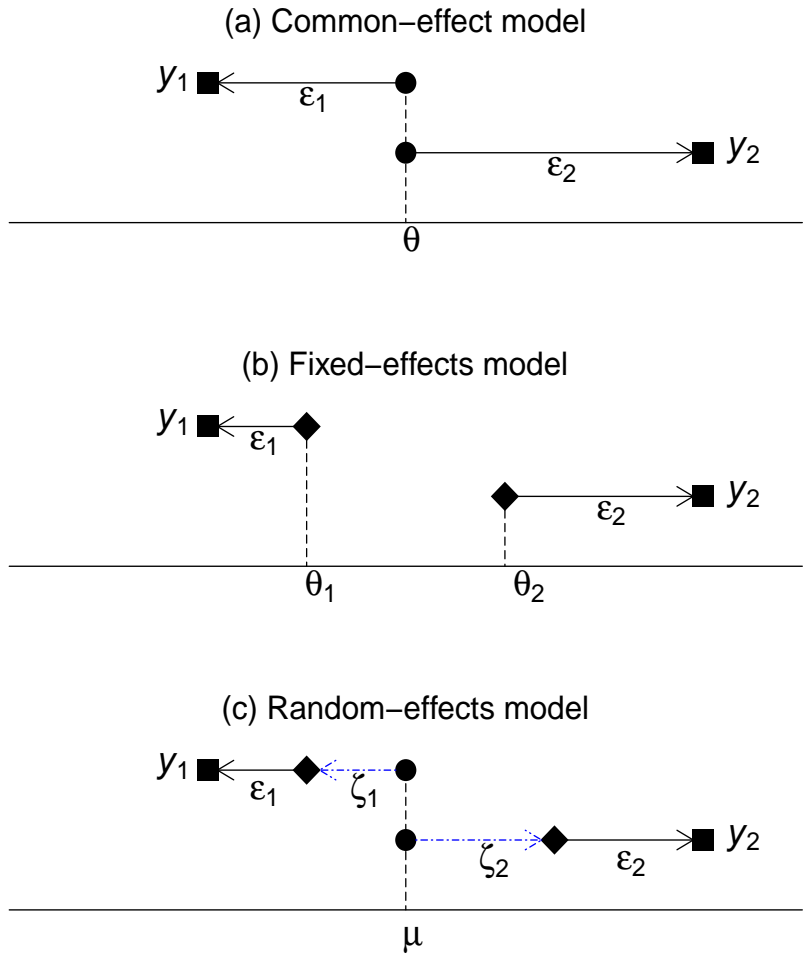}
	\caption{Schematics of the three models for meta-analysis: the common-effect model, the fixed-effects model, and  the random-effects model. The squares represent the observed effect sizes, the diamonds represent the study-specific effect sizes, and the circles represent the summary effects.}
	\label{fig:threemodelassumption}
  \end{figure}

\vskip 12pt
\subsection{Common-effect model}
\vspace{-1em}
\noindent
The common-effect model assumes that all the studies share a \textit{common effect}, for which we denote it by $\theta$.
In other words, as shown in Figure \ref{fig:threemodelassumption}(a), the differences in the observed effect sizes are all subject to sampling errors.
Mathematically, we can express the common-effect model as
\begin{equation} \label{common-effect model}
y_i=\theta+\varepsilon_{i},\ \ \  \varepsilon_{i} \overset{\rm ind}{\sim} N(0,\sigma_i^2),
\end{equation}
where $y_i$ are the observed effect sizes, $\theta$ is the \textit{common effect}, $\varepsilon_{i}$ are the random errors for studies $i=1, \dots, k$, and $k$ is the total number of studies.
In addition, the random errors $\varepsilon_i$ are assumed to be normally distributed with zero mean and variances $\sigma_i^2>0$, where ``ind'' represents the random errors are independent of each other.
To avoid confusion with the ``fixed-effects'' model in Figure 1(b) or more specifically in Section 2.3, we refrain from using the fixed-effect model, but refer to model (\ref{common-effect model}) consistently as the common-effect model throughout the paper.

With  model (\ref{common-effect model}), the \textit{common effect} can be unbiasedly estimated as
\begin{equation*}
\hat{\theta}=\frac{\sum_{i=1}^{k}w_i^{\mbox{\tiny C}} y_i}{\sum_{i=1}^{k}w_i^{\mbox{\tiny C}}},
\end{equation*}
where $w_i^{\mbox{\tiny C}}=1/\sigma_i^2$ are the study-specific weights that are derived by minimizing the mean squared error (MSE) of $\hat{\theta}$ (\citealp{laird1990some}).
For ease of reference, we have also provided a detailed derivation of the study-specific weights in Appendix A.
A study with a large sample size will more likely yield a small study-specific variance, and consequently, a large weight will be assigned to that study to quantify the level of reliability.

Further by the normality assumption on the random errors, $\hat{\theta}$ follows a normal distribution with mean $\theta$ and variance $1/\sum_{i=1}^{k}w_i^{\mbox{\tiny C}}$; and accordingly, a $95\%$ confidence interval (CI) of $\theta$ is $[\hat{\theta}- 1.96 /\sqrt{\sum_{i=1}^{k}w_i^{\mbox{\tiny C}}}, \hat{\theta}+1.96/\sqrt{\sum_{i=1}^{k}w_i^{\mbox{\tiny C}}}]$.
Note that the estimates of the within-study variances are treated as true variances to compute the \textit{common effect} and its CI.

\vskip 12pt
\subsection{Random-effects model}
\vspace{-1em}
\noindent
In many studies, the differences between the observed effect sizes can be relatively large such that they are not able to be explained fully by sampling errors.
Such differences are well known as the heterogeneity of the studies.
The heterogeneity may arise from various aspects, including, for example, the different designs in experiments or the different covariants in the model regression (\citealp{brockwell2001comparison}).
When such a scenario occurs, a \textit{common effect} for all the studies is unlikely to be true; and consequently, if model (\ref{common-effect model}) is still applied, it often yields underestimated standard errors of the estimated common effects and confidence intervals with insufficient coverage.

To account for the heterogeneity of the studies, as shown in Figure \ref{fig:threemodelassumption}(c), a random-effects model is often considered by assuming the study-specific effect sizes also follow an underlying distribution, e.g., a normal distribution.
To be more specific, the random-effects model can be expressed as
\begin{equation} \label{random-effects model}
y_i=\mu+\zeta_i+\varepsilon_{i}, \ \ \
\zeta_{i} \overset{\rm i.i.d.}{\sim} N(0,\tau^2), \ \ \
\varepsilon_{i} \overset{\rm ind}{\sim} N(0,\sigma_i^2),
\end{equation}
where $y_i$ are the observed effect sizes, $\mu$ is the \textit{mean effect},  $\zeta_i$ are the deviations of the study-specific effect sizes from the \textit{mean effect}, and $\varepsilon_{i}$ are the random errors.
We further assume that the deviations $\zeta_{i}$ are independent and identically distributed (i.i.d.) from $N(0,\tau^2)$, the random errors $\varepsilon_{i}$ are the same as in model (\ref{common-effect model}), and they are independent of each other.
In addition, we refer to $\tau^2$ as the between-study variance and $\sigma_i^2$ as the within-study variances.

With model (\ref{random-effects model}), if $\tau^2$ is known, the \textit{mean effect} can be unbiasedly estimated as
\begin{equation*}
\hat{\mu} =\frac{\sum_{i=1}^{k}w_i^{\mbox{\tiny R}}y_i}{\sum_{i=1}^{k}w_i^{\mbox{\tiny R}}},
\end{equation*}
where $w_i^{\mbox{\tiny R}}=1/(\sigma_i^2+\tau^2)$ are the study-specific weights that are derived by minimizing the MSE of $\hat{\mu}$ (\citealp{laird1990some}).
For a detailed derivation of the study-specific weights, please refer to Appendix B.
For the special case  when $\tau^2=0$, the random-effects model reduces to the common-effect model.
Further by the normality assumptions on both the effect sizes and the random errors, we have $\hat{\mu}$ following a normal distribution with mean $\mu$ and variance $1/\sum_{i=1}^{k}w_i^{\mbox{\tiny R}}$; and accordingly,  a $95\%$ CI of $\mu$ is $[\hat{\mu}-1.96 /\sqrt{\sum_{i=1}^{k}w_i^{\mbox{\tiny R}}}, \hat{\mu}+1.96/\sqrt{\sum_{i=1}^{k}w_i^{\mbox{\tiny R}}}]$.
Note that the between-study variance $\tau^2$ is unknown and needs to be estimated from the data.
In particular, there exists many methods for estimating $\tau^2$ in the literature including the moments estimator, the likelihood-based estimate, and the empirical Bayes estimate.
For a detailed review and comparison of these existing estimators, one may refer to  \cite{Sidik2010A}, \cite{veroniki2016methods}, and the references therein.
When few studies are included in a meta-analysis, a reliable estimate of the between-study variance is hard to achieve for most existing methods including the \cite{paule1982consensus} method and the \cite{dersimonian1986meta} method.
As illustrated in Section 4, the resulting CI may be too wide to be useful in practice.

\vskip 12pt
\begin{figure}
	\centering
	\includegraphics[width=1\linewidth, height=0.6\textheight]{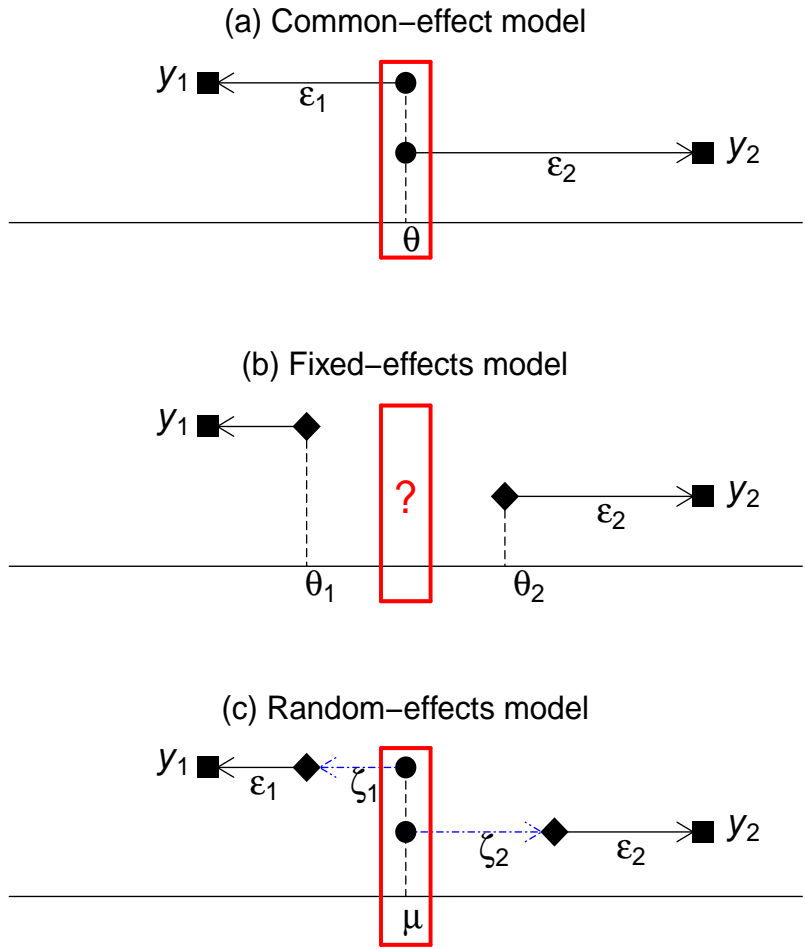}
	\caption{Schematics of the three models for meta-analysis: the common-effect model, the fixed-effects model, and the random-effects model. Unlike the common-effect model and the random-effects model, the fixed-effects model does not exist a clear position to place the circles for the summary effect.}
	\label{fig:dis-threemodelassumption}
\end{figure}

\subsection{Fixed-effects model}
\vspace{-1em}
\noindent
In the fixed-effects model, the effect size of each study is assumed to be a fixed but unequal quantity (\citealp{ramsey2012statistical}).
That is, we neither require the study-specific effect sizes to be all the same as in model (\ref{common-effect model}), nor assume that they follow a certain distribution as in model (\ref{random-effects model}).
Under these specifications, we can present  the fixed-effects model as
\begin{equation}\label{fixed-effects model}
y_i=\theta_i + \varepsilon_{i},   \    \      \
\varepsilon_{i} \overset{\rm ind}{\sim} N(0,\sigma_i^2),
\end{equation}
where $y_i$ are the observed effect sizes, $\theta_i$ are the study-specific effect sizes, and $\varepsilon_{i}$ are independent random errors  that are normally distributed with zero mean and variance $\sigma_i^2 > 0$.
Model (3) is referred to as a fixed-effects model since $\theta_{i}$ are deterministic rather than random variables as in model (2).

A major limitation of the fixed-effects model is that it does not provide a clear summary effect as the \textit{common effect} in model (\ref{common-effect model}) or the \textit{mean effect} in model (\ref{random-effects model}).
In other words, there does not exist a  clear position to place the circles for the summary effect, as shown in Figure \ref{fig:dis-threemodelassumption}.
Two summary effects have been proposed in the literature.
One is the \textit{unweighted average effect} as defined in \cite{laird1990some} and \cite{bender2018methods}, and the other is the \textit{weighted average effect} as defined in \cite{rice2018re}.

\emph{Unweighted Average Effect:} For the fixed-effects model, by treating each study as equally important,  one natural summary effect is
\begin{equation}\label{UAE}
\phi_u={1\over k}\sum_{i=1}^{k}\theta_i.
\end{equation}
We refer to this summary effect as the {\it unweighted average effect}.
Following this definition, \cite{laird1990some} proposed to estimate $\phi_u$ by
\begin{equation}\label{UAE-U}
\hat{\phi}_u={1\over k}\sum_{i=1}^{k}y_i.
\end{equation}
Further by the normality assumption in model (\ref{fixed-effects model}), $\hat{\phi}_u$ follows a normal distribution with mean $\sum_{i=1}^{k}\theta_i/k$ and variance $\sum_{i=1}^{k}\sigma_i^2 /k^2$, and a $95\%$ CI of $\phi_u$ is $[\hat{\phi}_u  - 1.96 \sqrt{\sum_{i=1}^{k}\sigma_i^2}/k, \hat{\phi}_u  + 1.96 \sqrt{\sum_{i=1}^{k}\sigma_i^2}/k]$.

\emph{Weighted Average Effect:} Allowing the weights to depend on the within-study variances and the sample sizes, \cite{rice2018re} proposed the \textit{weighted average effect} as
\begin{equation*}\label{WAE}
\phi_w=\frac{\sum_{i=1}^{k}\eta_i \gamma_i \theta_i}{\sum_{i=1}^{k}\eta_i \gamma_i} =\FS{\sum_{i=1}^{k} \FS{n_i}{\sum_{j=1}^{k}n_j} \FS{1}{n_i\sigma_i^2} \theta_i }{\sum_{i=1}^{k}\FS{n_i}{\sum_{j=1}^{k}n_j} \FS{1}{n_i\sigma_i^2}}=\FS{\sum_{i=1}^{k}\sigma_i^{-2}\theta_{i}}{\sum_{i=1}^{k}\sigma_i^{-2}},
\end{equation*}
where  $n_i$ are the sample sizes, $\eta_i={n_i}/{\sum_{j=1}^{k}n_j}$ are the  proportions associated with each individual  study, and $\gamma_i={1}/({n_i\sigma_i^2})$ reflect an average rate of information over the observations from the $i$th study.
They further proposed to estimate  the \textit{weighted average effect}  by
\begin{equation}\label{Rice-estimate}
\hat{\phi}_w= \frac{\sum_{i=1}^{k}\sigma_i^{-2}y_i}{\sum_{i=1}^{k}\sigma_i^{-2}}.
\end{equation}
Note that $\hat{\phi}_w$ is the same as the unbiased estimator of the \textit{common effect} $\theta$ in the common-effect model.
Under the normality assumption in model (\ref{fixed-effects model}), $\hat{\phi}_w$ follows a normal distribution with mean $\sum_{i=1}^{k}\theta_{i}\sigma_i^{-2}/\sum_{i=1}^{k}\sigma_i^{-2}$ and variance $1/\sum_{i=1}^{k}\sigma_i^{-2}$,
and a $95\%$ CI of $\phi_w$ is $[\hat{\phi}_w  - 1.96 /\sqrt{\sum_{i=1}^{k}\sigma_i^{-2}}, \hat{\phi}_w  + 1.96 /\sqrt{\sum_{i=1}^{k}\sigma_i^{-2}}]$.

\vskip 12pt
\section{Optimal estimation for the fixed-effects model}
\vspace{-1em}
\noindent
For better application of the fixed-effects model in meta-analysis with few studies, we propose an optimal estimator for the \textit{unweighted average effect} in this section.
We also assess the estimation accuracy of the proposed estimator through numerical studies.

\vskip 12pt
\subsection{Optimal estimator of the unweighted average effect}
\vspace{-1em}
\noindent
Note that  the \textit{common effect estimator} $\hat{\theta}$ and the \textit{mean effect estimator} $\hat{\mu}$ are both weighted averages of the observed effect sizes, in which the weights are assigned by the inverse-variance method.
Following the same logic, we propose to estimate the {\it unweighted average effect} $\phi_u$ in (\ref{UAE}) by
\begin{equation}\label{UAE-W}
\tilde{\phi}_u =\frac{\sum_{i=1}^{k}w_i^{\mbox{\tiny F}} y_i}{\sum_{i=1}^{k}w_i^{\mbox{\tiny F}}},
\end{equation}
where $w_i^{\mbox{\tiny F}}\geq0$ are the weights assigned to each individual study.
In the special case when $w_i^{\mbox{\tiny F}}=1/k$ for all $i=1,\dots,k$, our new estimator $\tilde{\phi}_u$ will be the same as the unweighted estimator $\hat{\phi}_u$ in (\ref{UAE-U}).
It is worth nothing that, even though $\hat\phi_u$ is an unbiased estimator, it does not guarantee to be the optimal estimator for $\phi_u$.

To derive the optimal weights in (\ref{UAE-W}), we choose to minimize the MSE  of $\tilde{\phi}_u$ as follows:
$$\text{MSE}(\tilde{\phi}_u ) = E\left(\frac{\sum_{i=1}^{k}w_i^{\mbox{\tiny F}} y_i}{\sum_{i=1}^{k}w_i^{\mbox{\tiny F}}}-\phi_u\right)^2.$$
We have shown in Appendix C that, when
\begin{equation}\label{assumption}
1+\sum_{j=1}^{k}\frac{(\theta_j-\theta_i)(\theta_j-\phi_u)}{\sigma_j^2} > 0,
\end{equation}
for $i =1,\dots,k$, there exists a unique positive solution for the optimal weights as
\begin{equation}\label{UAE-W2}
{w_i^{\mbox{\tiny F}}= \frac{1}{\sigma_i^2} \left(1+\sum_{j=1}^{k}\frac{(\theta_j-\theta_i)(\theta_j-\phi_u)}{\sigma_j^2}\right)}.
\end{equation}
Note that assumption (\ref{assumption}) is to guarantee the weights to be positive.
When the assumption does not hold, there is no explicit solution for the optimal weights.
In such settings, the computational methods such as interior point and conjugate gradient can be used to find the numerical solutions.
When $k=2$,  assumption (\ref{assumption}) always holds  and we have
\begin{equation*}
\left\{
\begin{array}{lr}
{w_1^{\mbox{\tiny F}}}=\FS{1}{\sigma_1^2}+\frac{(\theta_1-\theta_2)^2}{2{\sigma}_1^2{\sigma}_2^2}, &  \\
&  \\
{w_2^{\mbox{\tiny F}}}=\FS{1}{\sigma_2^2}+\frac{(\theta_1-\theta_2)^2}{2{\sigma}_1^2{\sigma}_2^2}. &
\end{array}
\right.
\end{equation*}
\vskip 3pt
\noindent
When $k=3$, if assumption (\ref{assumption}) holds, the optimal weights can also be expressed as
\begin{equation*}
\left\{
\begin{array}{lr}
w_1^{\mbox{\tiny F}}=\FS{1}{{\sigma}_1^2}+\frac{(\theta_1-\theta_2)(\theta_1+\theta_3-2\theta_2)}{3{\sigma}_1^2{\sigma}_2^2}+\frac{(\theta_1-\theta_3)(\theta_1+\theta_2-2\theta_3)}{3{\sigma}_1^2{\sigma}_3^2}, &  \\
&  \\
w_2^{\mbox{\tiny F}}=\FS{1}{{\sigma}_2^2}+\frac{(\theta_2-\theta_1)(\theta_2+\theta_3-2\theta_1)}{3{\sigma}_1^2{\sigma}_2^2}+\frac{(\theta_2-\theta_3)(\theta_1+\theta_2-2\theta_3)}{3{\sigma}_2^2{\sigma}_3^2}, &  \\
&  \\
w_3^{\mbox{\tiny F}}=\FS{1}{{\sigma}_3^2}+\frac{(\theta_3-\theta_1)(\theta_2+\theta_3-2\theta_1)}{3{\sigma}_1^2{\sigma}_3^2}+\frac{(\theta_3-\theta_2)(\theta_1+\theta_3-2\theta_2)}{3{\sigma}_2^2{\sigma}_3^2}. &
\end{array}
\right.
\end{equation*}
\vskip 3pt
\noindent

Compared to the  unbiased estimator $\hat{\phi}_u$ in (\ref{UAE-U}), our new estimator $\tilde{\phi}_u$ fully utilizes the information across the studies, and hence provides a smaller MSE.
Specifically, the bias of our new estimator is
$$\text{Bias}(\tilde{\phi}_u)= \frac{\sum_{i=1}^{k}w_i^{\mbox{\tiny F}} \theta_i}{\sum_{i=1}^{k}w_i^{\mbox{\tiny F}}}-\frac{\sum_{i=1}^{k}\theta_{i}}{k}.$$
When $\theta_i = \theta$ for all $i=1,\dots,k$, it follows that $\text{Bias}(\tilde{\phi}_u)=0$ and $w_i^{\mbox{\tiny F}}=1/\sigma_i^2$ which are exactly the same as $w_i^{\mbox{\tiny C}}$ in Section 2.1.
That is, when the fixed-effects model reduces to the common-effect model, our proposed estimator will be exactly the same as the \textit{common effect estimator} $\hat{\theta}$ in the common-effect model.
Note that $\tilde{\phi}_u$ is derived with minimizing the MSE,
which means $\text{MSE}(\hat{\phi}_u)-\text{MSE}(\tilde{\phi}_u)\geq 0.$
This leads to
$$\text{Var}(\hat{\phi}_u)-\text{Var}(\tilde{\phi}_u)=\text{MSE}(\hat{\phi}_u)-\text{MSE}(\tilde{\phi}_u)+ \text{Bias}(\tilde{\phi}_u)^2\geq 0,$$
where the equality holds only if  $\sigma_{1}^2=\dots=\sigma_{k}^2$, that is, $w_i^{\mbox{\tiny F}}/\sum_{j=1}^{k}w_j^{\mbox{\tiny F}}=1/k$ so that $\tilde{\phi}_u$ is the same as $\hat{\phi}_u$.
This shows that our new estimator always has a smaller variance than the existing estimator.
Taking $k=2$ as an example, the difference between the two variances can be expressed as $(\sigma_1^2-\sigma_2^2)^2[\sigma_1^2+\sigma_2^2+2(\theta_1-\theta_2)^2]/\{[\sigma_1^2+\sigma_2^2+(\theta_1-\theta_2)^2]^2\}$
which is non-negative with the equality holding only when $\sigma_1^2=\sigma_2^2$.

From another perspective, we note that
$\tilde{\phi}_u=\alpha \hat{\phi}_w + (1-\alpha) \hat{\phi}_u$
where $\alpha=\sum_{i=1}^k \sigma_i^{-2}/\sum_{i=1}^k w_i^{\mbox{\tiny F}}$.
Therefore, the optimal estimator is a weighted average of the \textit{unweighted average effect estimator}  $\hat{\phi}_u$ by \cite{laird1990some} and the \textit{weighted average effect estimator}  $\hat{\phi}_w$ by \cite{rice2018re}.
When all $\theta_i$ are the same, it reduces to $\hat{\phi}_w$ in (\ref{Rice-estimate}).
Finally, under the assumptions of the fixed-effects model,
$\tilde{\phi}_u $ follows a normal distribution with mean $\sum_{i=1}^{k}w_i^{\mbox{\tiny F}} \theta_i/\sum_{i=1}^{k}w_i^{\mbox{\tiny F}}$ and variance $\sum_{i=1}^{k}({w_i^{\mbox{\tiny F}}}\sigma_i)^2/(\sum_{i=1}^{k}w_i^{\mbox{\tiny F}})^2$,
and a $95\%$ CI of $\phi_u$ is
$[\tilde{\phi}_u - 1.96 \sqrt{\sum_{i=1}^{k}({{w_i^{\mbox{\tiny F}}}\sigma_i})^2}/\sum_{i=1}^{k}w_i^{\mbox{\tiny F}},
\tilde{\phi}_u +1.96 \sqrt{\sum_{i=1}^{k}({w_i^{\mbox{\tiny F}}}\sigma_i)^2}/\sum_{i=1}^{k}w_i^{\mbox{\tiny F}}]$.

\vskip 12pt
\subsection{Numerical comparison}
\begin{figure}[htbp]
	\centering
	\includegraphics[width=1\textwidth,height=0.5\textheight]{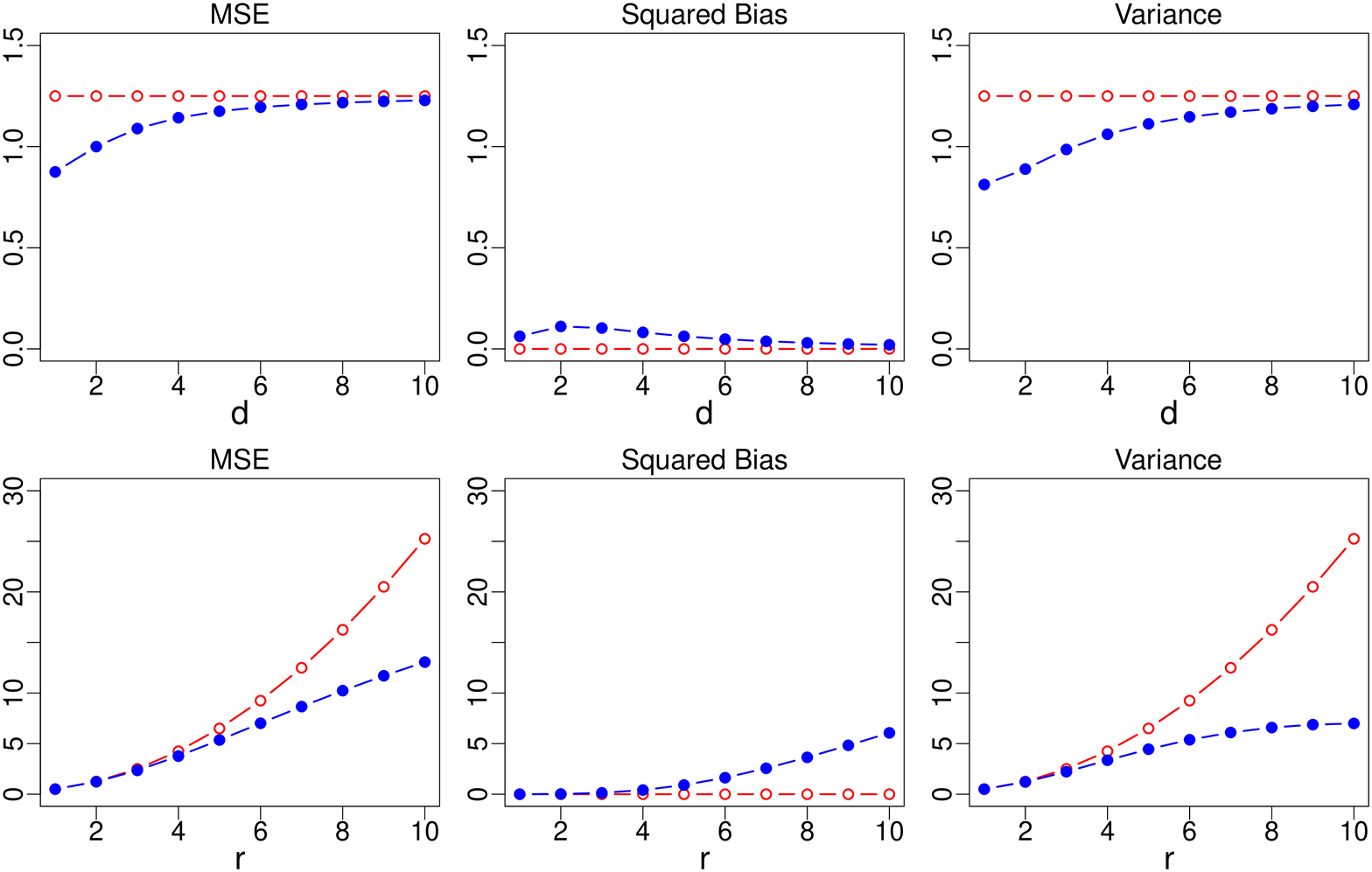}
	\caption{Comparison of the two estimators for the \textit{unweighted average effect} with $k=2$. The red lines with empty circles represent the results of the unbiased estimator $\hat{\phi}_u$, and the blue lines with solid circles represent the results of our new estimator $\tilde{\phi}_u$. The top three plots represent the results along with the range of $d$, and the bottom three plots represent the results along with the range of $r$.}
	\label{fig:numerical-2}
\end{figure}

\begin{figure}[htbp]
	\centering
	\includegraphics[width=1\textwidth,height=0.5\textheight]{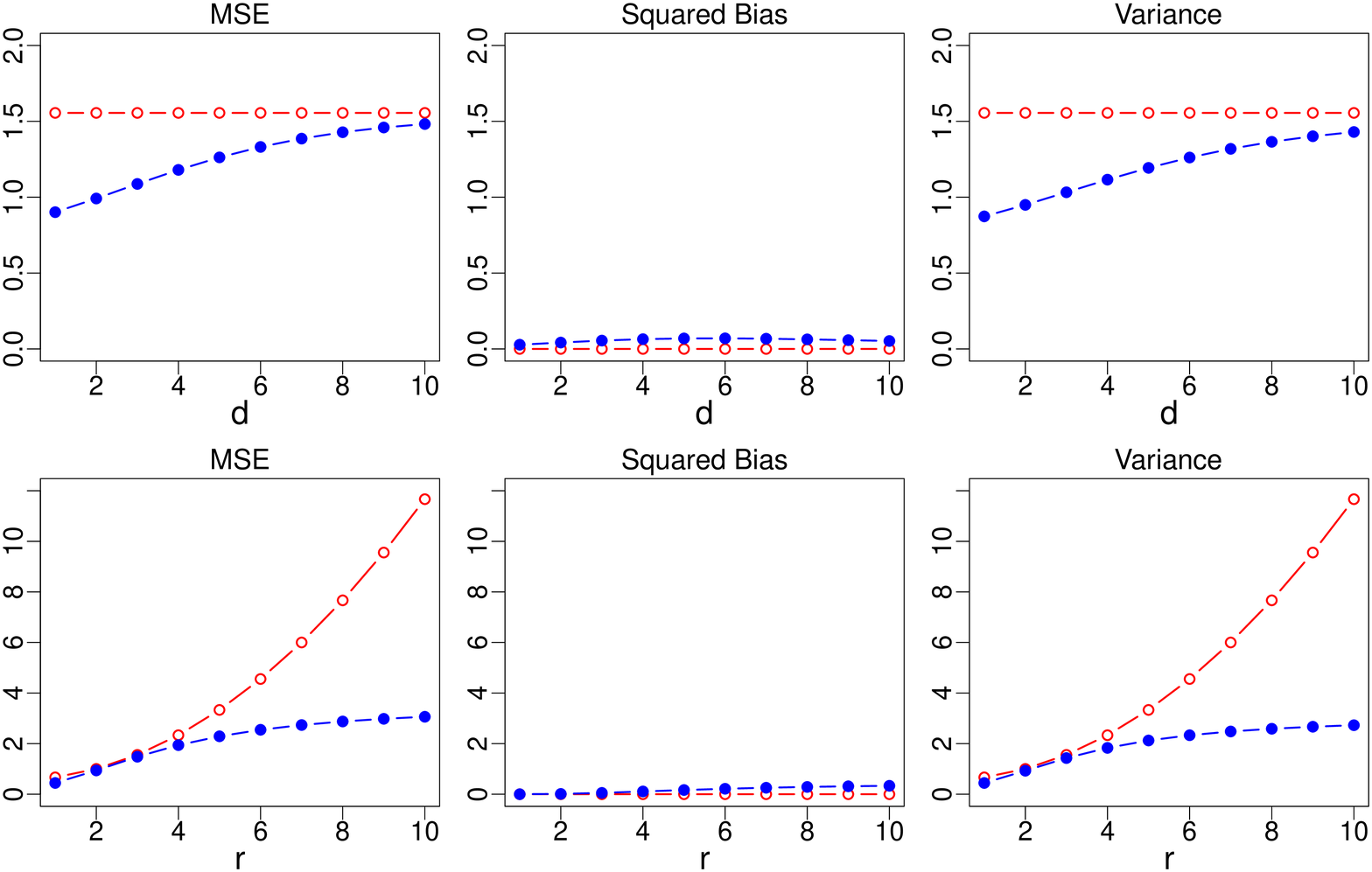}
	\caption{Comparison of the two estimators for the \textit{unweighted average effect} with $k=3$. The red lines with empty circles represent the results of the unbiased estimator $\hat{\phi}_u$, and the blue lines with solid circles represent the results of our new estimator $\tilde{\phi}_u$. The top three plots represent the results along with the range of $d$, and the bottom three plots represent the results along with the range of $r$.}
	\label{fig:numerical-3}
\end{figure}

\vspace{-1em}
\noindent
Since a half or more of meta-analyses include only two or three studies, we therefore conduct numerical studies to evaluate the performance of our new estimator for  $k=2$ and 3.
The unbiased estimator $\hat{\phi}_u$ is also considered for comparison to explore how much improvement that our new estimator can achieve.
To visualize the numerical results, we compute the MSE, the squared bias, and the variance of the two estimators, and plot them in Figure \ref{fig:numerical-2} for $k=2$ and in Figure \ref{fig:numerical-3} for $k=3$.

In the first study with $k=2$,
we consider the fixed-effects model in (\ref{fixed-effects model}) with several different settings of  $\theta_i$ and $\sigma_i$.
Specifically for the top three plots in Figure \ref{fig:numerical-2}, we consider $\sigma_{1}=1$, $\sigma_{2}=2$, $\theta_1=0$, and $\theta_2-\theta_1=d$ with the difference $d$ ranging from 0 to 10; while for the bottom three plots, we consider $\theta_1=-5$, $\theta_2=5$, $\sigma_{1}=1$, and $\sigma_{2}/\sigma_{1}=r$ with the ratio $r$ ranging from 1 to 10.
Our second study is with $k=3$.
For the top three plots in Figure \ref{fig:numerical-3}, we consider  $\sigma_{1}=1$, $\sigma_{2}=2$, $\sigma_{3}=3$, $\theta_1=0$, $\theta_2=5$, and $\theta_3-\theta_1=d$ with the difference $d$ ranging from 0 to 10;
while for the bottom three plots, we consider  $\theta_1=-10$, $\theta_2=0$, $\theta_3=10$, $\sigma_{1}=1$, $\sigma_{2}=2$, and $\sigma_{3}/\sigma_{1}=r$ with the ratio $r$ ranging from 1 to 10.

From the numerical results, it is evident that our optimal estimator has a smaller MSE than the unbiased estimator in all settings for $k=2$ and $k=3$. It is also evident that the MSE as the sum of the squared bias and the variance is mainly contributed by the variance of the estimator.
The improvement increases as the difference between means $\theta_i$ decreases or the ratio between variances $\sigma_i$ increases.
Note that we have also conducted numerical studies for many other settings (not shown to save space), and the comparison results remain similar as those in Figures \ref{fig:numerical-2} and \ref{fig:numerical-3}. To conclude, in comparison with  the unbiased estimator, our new estimator sacrifices a small bias for a significant reduction in variance of the estimator so that the MSE decreases.

\vskip 12pt
\section{Applications to meta-analysis with few studies}
\vspace{-1em}
\noindent
In this section, we  apply the fixed-effects model to four real  examples in clinical trails, including studies with both continuous and binary outcomes, and compare them with the common-effect model and the random-effects model.
We compute both the optimal estimator and the \citeauthor{laird1990some} (L-M) estimator for the  \textit{unweighted average effect}  in the fixed-effects model.
\vskip 12pt
\subsection{Meta-analysis with two studies}
\begin{figure}[htbp]
	\centering
	\includegraphics[width=1\textwidth,height=0.4\textheight]{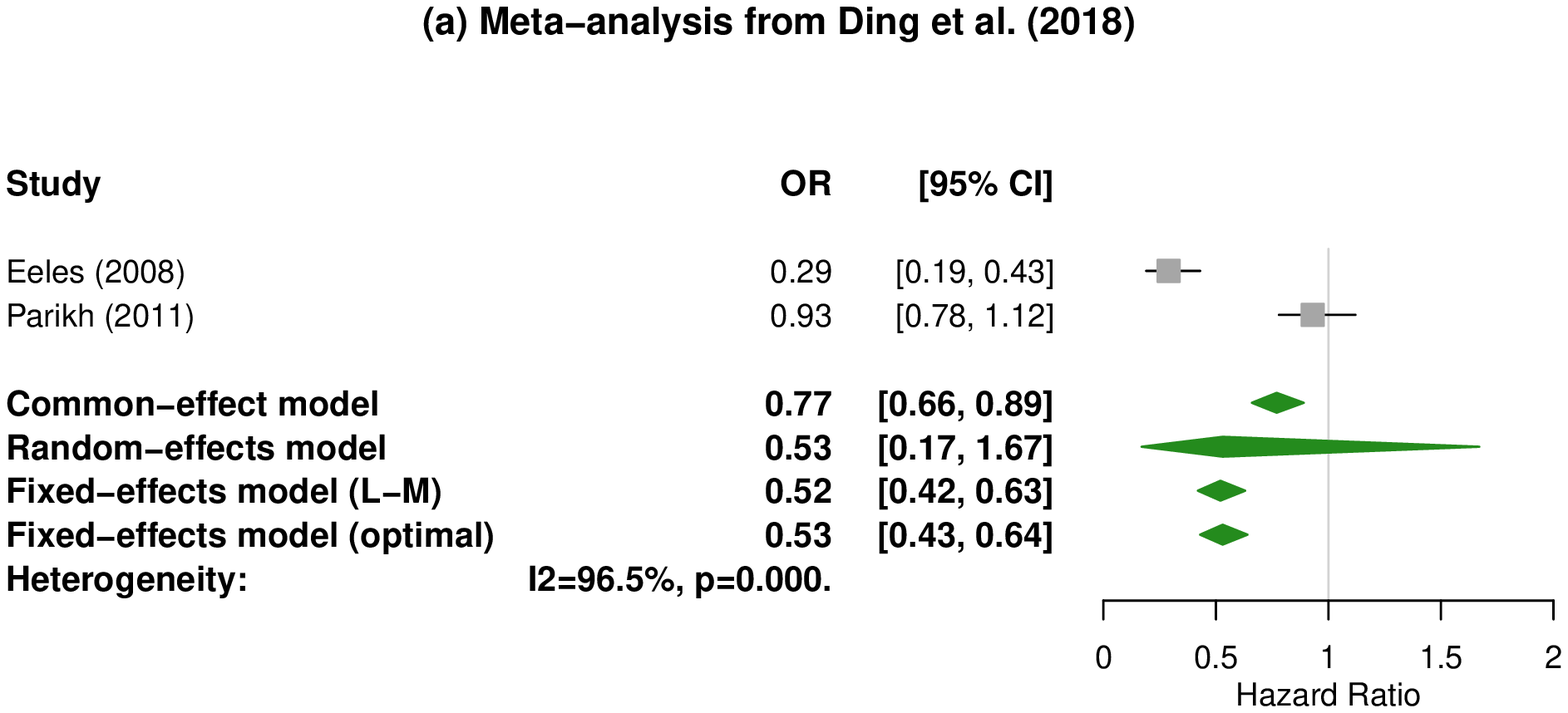}
	\includegraphics[width=1\textwidth,height=0.4\textheight]{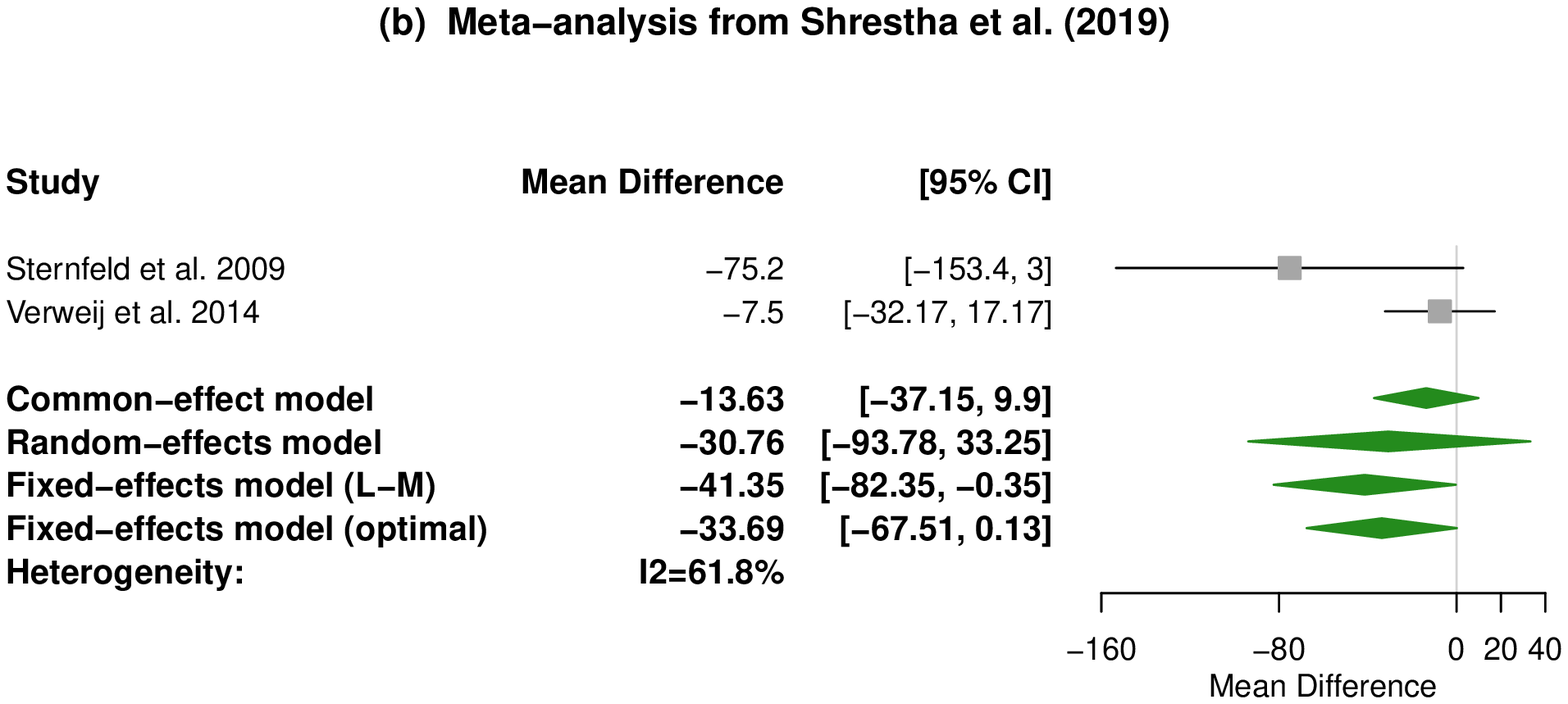}
	\caption{Meta-analyses with exactly two studies, where (a) represents the association between the genetic variant rs1058205 in kallikrein 3 (KLK3) and prostate cancer risk, and (b) represents the association between asthma and periodontal disease in adults.}
	\label{fig:real_example_2}
\end{figure}

\vspace{-1em}
\noindent
The first example is with \cite{ding2018association}, who reported a systematic review about the association between three genetic variants (rs1058205, rs2735839, and rs266882) in kallikrein 3 (KLK3) and prostate cancer risk, with the observed effect sizes being the odds ratios (OR).
The genotypes reflected the disease include CC, CT, and TT.
In a subgroup meta-analysis, only two studies are included as shown in Figure \ref{fig:real_example_2}(a).
The first study indicates a significant association between the rs1058205 TT and CC (OR = 0.29, CI = 0.19 to 0.43), whereas the second study does not give a significant result (OR = 0.93, CI = 0.78 to 1.12).
Combining the two studies, the common-effect model gives a significant result for the \textit{common effect} (OR = 0.77, CI = 0.66 to 0.89).
Nevertheless, with the large heterogeneity between the two studies ($I^2=96.5\%$, $p<0.001$), the assumption of a \textit{common effect} may not hold so that the meta-analytic results from the common-effect model may be misleading.
On the other hand, the random-effects model does not provide a significant result for the \textit{mean effect} with a wide CI (OR = 0.53, CI = 0.17 to 1.67).
In contrast, different from the above two models,  the fixed-effects model does not involve an unstable estimate of the between-study variance, and meanwhile it does not impose the unrealistic assumption of a \textit{common effect}.
Specifically, it provides significant results for the \textit{unweighted average effect} (the L-M estimator: OR = 0.52, CI = 0.42 to 0.63; the optimal estimator: OR = 0.53, CI = 0.43 to 0.64), which is very close to the \textit{mean effect} in the random-effects model (OR = 0.53); and more importantly, it yields a narrower CI for the \textit{unweighted average effect} which is about the same length as that for the  the \textit{common effect} in the common-effect model.

The second example  investigated the effectiveness of interventions for reducing nonoccupational sedentary behavior in adults and older adults (\citealp{shrestha2019effectiveness}).
As shown in  Figure \ref{fig:real_example_2}(b),  the observed effect sizes are the mean differences (MD).
Neither study showed a significant difference between the two groups (study 1: MD = -75.2, CI = -153.4 to 3; study 2: MD = -7.5, CI = -32.17 to 17.17).
Combining the two studies, the common-effect model gives a non-significant result for the \textit{common effect} (MD = -13.63, CI = -37.15 to 9.9).
The random-effects model also does not have a significant result for the \textit{mean effect} (MD = -30.76, CI = -93.78 to 33.25).
To summarize, neither the common-effect model nor the random-effects model reports a significant result.
Nevertheless, we note that the fixed-effects model with the L-M estimator reports a significant result (MD = -41.35, CI = -82.35 to -0.35), which may not be very reasonable.
Fortunately, the fixed-effects model with our optimal estimator leads to the same non-significant result as the common-effect model and the random-effects model (MD = -33.69, CI = -67.51 to 0.13), which is close to  the estimate of the \textit{mean effect} in the random-effects model.

\vskip 12pt
\subsection{Meta-analysis with three studies}
\begin{figure}[htbp]
	\centering
	\includegraphics[width=1\textwidth,height=0.4\textheight]{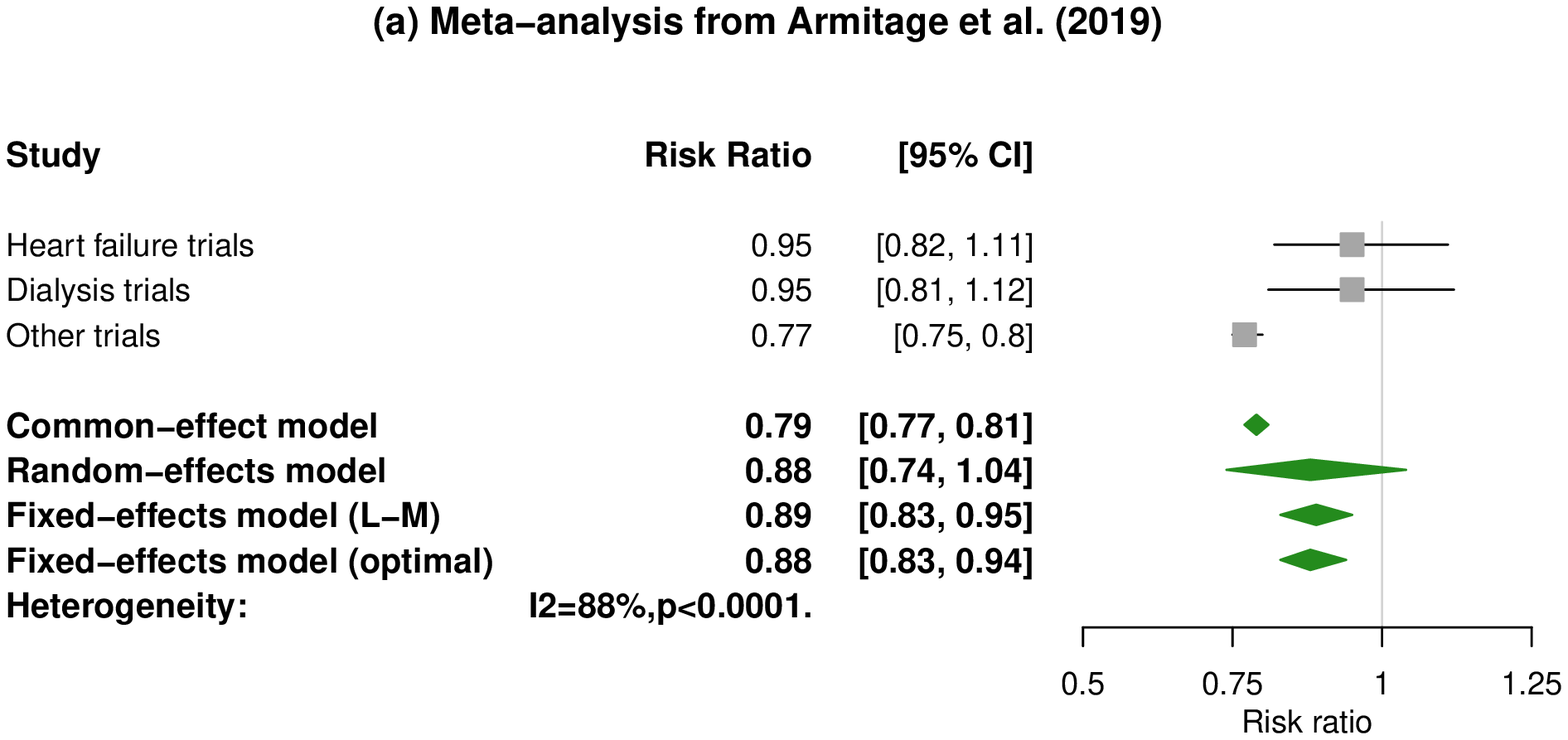}
	\includegraphics[width=1\textwidth,height=0.4\textheight]{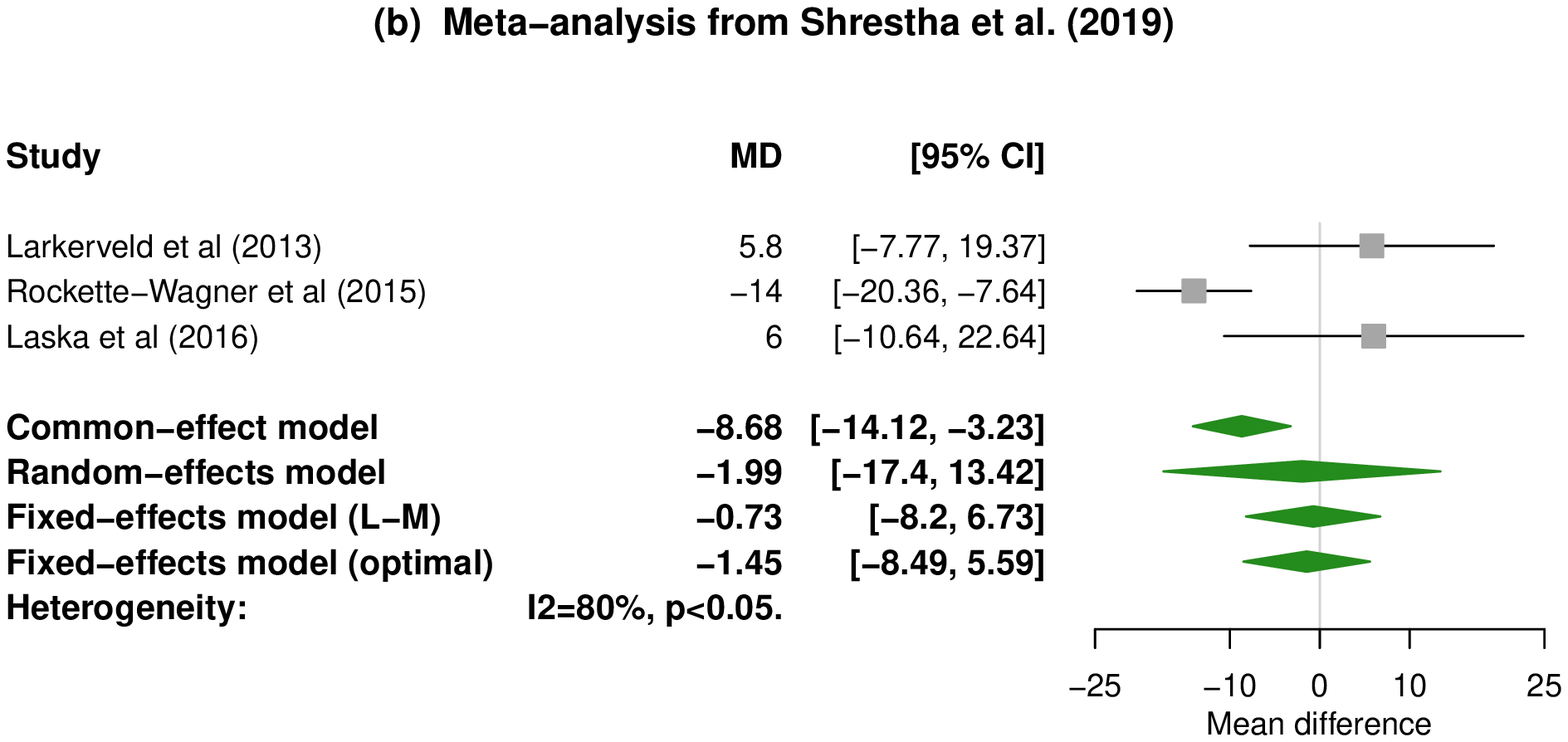}
	\caption{Meta-analyses with exactly three studies, where (a) represents the efficacy and safety of statin therapy in older people, and (b) represents the effectiveness of interventions for reducing nonoccupational sedentary behavior in adults and older adults.}
	\label{fig:real_example_3}
\end{figure}

\vspace{-1em}
\noindent
The first example is with \cite{armitage2019efficacy}.
The authors reported a systematic review about the efficacy and safety of statin therapy in older people, with the observed effect sizes being the  risk ratios (RR).
In a subgroup of effects on major vascular events per mmol/L reduction in LDL cholesterol, subdivided by the age larger than 75, only three studies are included for meta-analysis as shown in Figure \ref{fig:real_example_3}(a).
The first two studies both report non-significant results (study 1: RR = 0.95, CI = 0.82 to 1.11; study 2: RR = 0.95, CI = 0.81 to 1.12), and third study shows a significant result (RR = 0.77, CI = 0.75 to 0.80).
Combining the three studies, the common-effect model gives a significant result for the \textit{common effect} (RR = 0.79, CI = 0.77 to 0.81).
Nevertheless, with the large heterogeneity among the three studies ($I^2=88\%$, $p<0.001$), the meta-analytic results based on a \textit{common effect} assumption may be misleading in practice.
On the other hand,  the random-effects model does not show a significant result for the \textit{mean effect} with a wide CI (RR = 0.88, CI = 0.74 to 1.04).
Compared to the two models, the fixed-effects model provides significant results for the \textit{unweighted average effect} (the L-M estimator: RR = 0.89, CI = 0.83 to 0.95; the optimal estimator: RR = 0.88, CI = 0.83 to 0.94), which is very close to  the estimate of the \textit{mean effect} in the random-effects model; and both estimators yield narrower CIs for  the \textit{unweighted average effect} than that for  the \textit{mean effect} in the random-effects model.

The second example is extracted  from the systematic review  to investigate the effectiveness of interventions for reducing nonoccupational sedentary behavior in adults and older adults (\citealp{shrestha2019effectiveness}, Figure \ref{fig:real_example_3}(b)).
We address the outcomes in the subgroup of the continuous outcomes, which includes only three studies.
The observed effect sizes in this meta-analysis are the mean differences (MD).
Specifically, the first study reports a non-significant result (MD = 5.8, CI = -7.77 to 19.37), the second study also reports a significant result (MD = -14, CI = -20.36 to -7.64),
and the third study reports a non-significant result (MD = 6, CI = -10.64 to 22.64).
Combining the three studies, the common-effect model gives a significant result for the \textit{common effect} (MD = -8.68, CI = -14.12 to -3.23).
Nevertheless, with the large heterogeneity among the three studies ($I^2=80\%$, $p<0.005$), the meta-analytic results from the common-effect model may be misleading.
In contrast, the random-effects model does not offer a significant result for the \textit{mean effect} (MD = -1.99, CI = -17.4 to 13.42).
In this example, we note that  the fixed-effects model also show non-significant results by both estimators  (the L-M estimator: MD = -0.73, CI = -8.2 to 6.73; the optimal estimator: MD = -1.45, CI = -8.49 to 5.59).
Specifically, it provides a point estimate of the \textit{unweighted average effect} that is nearly the same as that for  the \textit{mean effect} in the random-effects model (SMD = -1.99), and meanwhile it produces a much narrower  CI for  the \textit{unweighted average effect}  than that for the \textit{mean effect} in the random-effects  model.
Note also that the optimal estimator has, once  again, a slightly narrower CI than the L-M estimator.

Based on the four real examples, we find that, based on different assumptions, the three different models may have different summary effect estimates and conclusions.
The common-effect model should be avoided in the presence of heterogeneity.
On the other hand, the CI for the \textit{mean effect} of the random-effects model is usually too wide to be useful when the number of studies is small.
The fixed-effects model tends to have estimates of \textit{unweighted average effect} close to the estimate of the \textit{mean effect} in the random-effects model with much narrower CIs.
Therefore, they provide a more meaningful summary for the effect sizes when the number of studies is small and the heterogeneity is non-negligible.

\vskip 12pt
\section{Conclusion}
\vspace{-1em}
\noindent
Two commonly used models for meta-analysis, the common-effect model and the random-effects model, both have their own limitations.
The common-effect model  may lead to misleading results when the heterogeneity among studies cannot be ignored.
The CI for the \textit{mean effect} based on the random-effects model may be too wide to be useful  when the number of studies is small.
These limitations are major obstacles in practice since a large portion of meta-analyses have small numbers of studies with non-negligible heterogeneity.
On the other hand, the fixed-effects model not only avoids the unreasonable assumption on a \textit{common effect} in the common-effect model when the heterogeneity exists, but also avoids the low accuracy of the between-study variance estimate in the random-effects model when there are only few studies.

In this paper, we demonstrated that the fixed-effects model can serve as an alternative approach for meta-analysis under these situations with conclusions limited to the observed studies.
We also proposed an optimal estimator for the \textit{unweighted average effect} in the fixed-effects model that fully utilizes the information across the studies.
With a small increase in bias but a large reduction in variance, the optimal estimator always has a smaller MSE than the existing estimator in the literature.
Also as shown in Figure \ref{fig:dis-threemodelassumption}, which quantity should be used as a summary effect in the fixed-effects model is not immediately clear.
Both the \textit{unweighted and weighted average effects} were proposed in the literature.
The \textit{unweighted average effect} treats each study as equally important and is easy to interpret (\citealp{laird1990some}; \citealp{bender2018methods}).
In contrast, the \textit{weighted average effect} allows researchers to assign different weights to different studies (\citealp{dominguez2018addressing}; \citealp{rice2018re}).
Which average effect to use for the fixed-effects model  may depend on different factors such as the goal of the meta-analysis and the sample sizes of different studies, and may require further research.

\vskip 12pt

\bibliographystyle{apalike2}
\urlstyle{same}
\bibliography{Fixed-effects_model}
\newpage
\noindent
\newpage
\centerline{\large {\bf Appendix A}}
\vskip 12pt
\noindent
For the common-effect model, we consider the weighted average as an estimator of $\theta$ as
$$\hat{\theta}=\sum_{i=1}^{k}w_i y_i,$$
where $\sum_{i=1}^{k}w_i=1$.
It is evident that $\hat{\theta}$ is an unbiased estimator of $\theta$.
Consequently, the variance of $\hat{\theta}$ will be the same as its MSE.
Then with the Lagrange term, we have
\begin{eqnarray*}
	f(w_1, \ldots, w_k,\lambda)
	&=& \text{MSE}(\hat{\theta}) - \lambda (\sum_{i=1}^{k}w_i-1)	\nonumber\\
	&=& E(\sum_{i=1}^{k}w_iy_i-\theta)^2 - \lambda (\sum_{i=1}^{k}w_i-1)	\nonumber\\
	&=& \sum_{i=1}^{k} {w_i}^2 \sigma_i^2 - \lambda (\sum_{i=1}^{k}w_i-1),
\end{eqnarray*}
where $\lambda$ is a Lagrange multiplier to enforce the constraint $\sum_{i=1}^{k}w_i-1=0$.
To derive the study-specific weights, we take the first derivatives and set them to zero,
$${\partial \over \partial w_i}f(w_1, \ldots, w_k,\lambda)=2\sigma_i^2 w_i^*- \lambda=0, ~~~~i=1,\dots ,k,$$
we have $w_i=\lambda/2\sigma_i^2$ for $i=1,\dots,k$.
Combining it with ${\partial \over \partial \lambda}f(w_1, \ldots, w_k,\lambda)=\sum_{i=1}^{k} w_i-1=0,$ we have $\lambda=2/(\sum_{i=1}^{k}1/\sigma_i^2)$ and $w_i=(1/\sigma_i^2)/(\sum_{i=1}^{k}1/\sigma_i^2)$.
Finally, we have
\begin{equation*}
\hat{\theta}=\frac{\sum_{i=1}^{k}w_i^{\mbox{\tiny C}} y_i}{\sum_{i=1}^{k}w_i^{\mbox{\tiny C}}},
\end{equation*}
where $w_i^{\mbox{\tiny C}}=1/\sigma_i^2$ for $i=1,\dots, k$.

\vskip 12pt
\vskip 12pt
\centerline{\large {\bf Appendix B}}
\vskip 12pt
\noindent
For the random-effects model, we also consider the weighted average as an estimator of $\mu$ as
$$\hat{\mu}=\sum_{i=1}^{k}w_i y_i,$$
where $\sum_{i=1}^{k}w_i=1$.
It is evident that $ \hat{\mu}$ is an unbiased estimator of $\mu$.
Consequently, the variance of $ \hat{\mu}$ will be the same as its MSE.
Then with the Lagrange term, we have
\begin{eqnarray*}
	f(w_1, \ldots, w_k,\lambda)
	&=& \text{MSE}(\hat{\mu}) - \lambda (\sum_{i=1}^{k}w_i-1)	\nonumber\\
	&=& E(\sum_{i=1}^{k}w_iy_i-\mu)^2 - \lambda (\sum_{i=1}^{k}w_i-1)	\nonumber\\
	&=& \sum_{i=1}^{k} {w_i}^2 (\sigma_i^2+\tau^2) - \lambda (\sum_{i=1}^{k}w_i-1),
\end{eqnarray*}
where $\lambda$ is a Lagrange multiplier to enforce the constraint $\sum_{i=1}^{k}w_i-1=0$.
To derive the study-specific weights, we take the first derivatives and set them to zero,
$${\partial \over \partial w_i}f(w_1, \ldots, w_k,\lambda)=2(\sigma_i^2+\tau^2) w_i- \lambda=0, ~~~~i=1,\dots, k,$$
we have $w_i=\lambda/2(\sigma_i^2+\tau^2)$ for $i=1,\dots,k$.
Combining it with ${\partial \over \partial \lambda}f(w_1, \ldots, w_k,\lambda)=\sum_{i=1}^{k} w_i-1=0,$ we have $\lambda=2/(\sum_{i=1}^{k}1/(\sigma_i^2+\tau^2))$.
Finally, we have
\begin{equation*}
\hat{\mu}=\frac{\sum_{i=1}^{k}w_i^{\mbox{\tiny R}} y_i}{\sum_{i=1}^{k}w_i^{\mbox{\tiny R}}},
\end{equation*}
where $w_i^{\mbox{\tiny R}}=1/(\sigma_i^2+\tau^2)$ for $i=1,\dots, k$.

\vskip 12pt
\vskip 12pt
\centerline{\large {\bf Appendix C}}
\vskip 12pt
\noindent
For the fixed-effects model, we consider the weighted average $\tilde{\phi}_u =\sum_{i=1}^{k}w_i y_i$ as an estimator of $\phi_u=\sum_{i=1}^{k}\theta_i/k$.
To solve the optimal weights, we have
\begin{equation}\label{Quadratic programming}
\text{min}~ f(w_1, \ldots, w_{k}),   ~~~\text{s.t.} ~~~ \sum_{i=1}^{k}w_i-1=0;~ w_j \geq 0,~ j=1,\dots, k,
\end{equation}
where
\begin{eqnarray*}
	f(w_1, \ldots, w_{k})
	&=& \text{MSE}(\tilde{\phi}_u )	\nonumber\\
	&=& E(\sum_{i=1}^{k}w_iy_i-\phi_u)^2 	\nonumber\\
	&=& \sum_{i=1}^{k} {w_i}^2\sigma_i^2+(\sum_{i=1}^{k}w_i \theta_i)^2 -2\sum_{i=1}^{k}w_i \phi_u \theta_i +\phi_u^2.
\end{eqnarray*}
Note that for any two different vectors, $\bfw=(w_1, \ldots, w_{k})$ and $\bfw^*=(w_1^*, \ldots, w_{k}^*)$, and for any $0<t<1$, we have
\begin{eqnarray*}
	&&	f((1-t)\bfw+t\bfw^*) -(1-t)f(\bfw) -tf(\bfw^*) \nonumber\\
	&=& t(t-1)\left(\sum_{i=1}^{k}\sigma_i^2({w_i-w_i^*)}^2
	+\left[\sum_{i=1}^{k}\theta_i(w_i-w_i^*)\right]^2\right)	\nonumber\\
	&<& 0,
\end{eqnarray*}
which means that $f(w_1, \ldots, w_{k})$ is a strictly convex function.
Thus the quadratic programming problem (\ref{Quadratic programming}) has a unique global minima for the feasible region $\{\bfw| \sum_{i=1}^{k}w_i-1=0;~ w_j \geq 0, j=1,\dots, k\}$;
and the sufficient and necessary condition of $\bfw$ being the global minima  is that $\bfw$ satisfies the Karush-Kuhn-Tucker (KKT) conditions as follows:
\begin{equation} \label{eqn2}
\begin{split}
{\partial \over \partial w_i}L(w_1,\dots, w_k, \lambda, \mu_1, \dots, \mu_k)&=0,    \\
\sum_{j=1}^{k}w_j-1&=0,    \\
w_i &\geq 0,    \\
\mu_i &\geq 0,    \\
\mu_iw_i&=0,
\end{split}
\end{equation}
for $i=1,\dots, k$.
The Lagrangian function $L(w_1,\dots, w_k, \lambda, \mu_1, \dots, \mu_k)$ is given as
\begin{eqnarray*}
	L(w_1,\dots, w_k, \lambda, \mu_1, \dots, \mu_k)
	&=& f(w_1, \ldots,w_k) +\lambda(1-\sum_{i=1}^{k}w_i)-\sum_{i=1}^k\mu_iw_i	\nonumber\\
	&=& \sum_{i=1}^{k} {w_i}^2\sigma_i^2+(\sum_{i=1}^{k}w_i \theta_i)^2 -2\sum_{i=1}^{k}w_i \phi_u \theta_i +\phi_u^2\nonumber\\
	&&+ ~\lambda (1-\sum_{i=1}^{k}w_i)-\sum_{i=1}^k\mu_iw_i,
\end{eqnarray*}
where $\lambda$ and $\mu_i$ are Lagrange multipliers to enforce the constraints.

When the quadratic programming problem (\ref{Quadratic programming}) has an interior solution, which means $w_i> 0$ and $\mu_i=0$ for  $i=1,\dots, k$,
there exists an explicit solution under the KKT conditions:
$$\lambda=\frac{2+2(\sum_{j=1}^{k}w_j\theta_j-\phi_u)\sum_{j=1}^{k} \frac{\theta_j}{\sigma_j^2}}{\sum_{j=1}^{k} \frac{1}{\sigma_j^2}},$$
and
$$w_i=a_i-a_ib_i\phi_u+a_ib_i\sum_{j=1}^{k}w_j\theta_j,$$
where $a_i=\sigma_i^{-2}/(\sum_{j=1}^{k} \sigma_{j}^{-2})$ and $b_i= \sum_{j=1}^{k}(\theta_j-\theta_i)\sigma_i^{-2}.$
For each $i$, we have one equation with $k$ variables for $i=1,\ldots, k$. Then we can solve all the $w_i$ with $k$ equations as
$$\bfA \bfw=\bfB,$$
where $\bfw=(w_1, \ldots, w_k)^T$, $\bfB=\left(a_1b_1\phi_u-a_1, \ldots, a_kb_k\phi_u-a_k\right)^T$, and
$$\bfA=\left(
\begin{array}{cccc}
a_1b_1\theta_1-1 & a_1b_1\theta_2 & \ldots & a_1b_1\theta_k \\
a_2b_2\theta_1& a_2b_2\theta_2-1 & \ldots & a_2b_2\theta_k \\
\vdots& \vdots & \ddots & \vdots \\
a_kb_k\theta_1& \ldots & a_kb_k\theta_{k-1} & a_kb_k\theta_k-1 \\
\end{array}
\right).$$
The unique solution is given as
\begin{eqnarray*}
	w_i
	&=& \frac{\frac{1}{\sigma_i^2} \left(1+\sum_{j=1}^{k}\frac{(\theta_j-\theta_i)(\theta_j-\phi_u)}{\sigma_j^2}\right)}{\sum_{j=1}^{k}\frac{1}{\sigma_j^2} \left(1+\sum_{l=1}^{k}\frac{(\theta_l-\theta_j)(\theta_l-\phi_u)}{\sigma_l^2}\right)}, \nonumber\\
\end{eqnarray*}
for $i=1, \dots, k$.
When $1+\sum_{j=1}^{k}(\theta_j-\theta_i)(\theta_j-\phi_u)/{\sigma_j^2} > 0$ for all $i=1, \dots, k$,
we  have
\begin{equation*}
\tilde{\phi}_u=\frac{\sum_{i=1}^{k}w_i^{\mbox{\tiny F}} y_i}{\sum_{i=1}^{k}w_i^{\mbox{\tiny F}}},
\end{equation*} 	
and the positive weights are
$${w_i^{\mbox{\tiny F}}= \frac{1}{\sigma_i^2} \left(1+\sum_{j=1}^{k}\frac{(\theta_j-\theta_i)(\theta_j-\phi_u)}{\sigma_j^2}\right)}.$$
Note that the inequalities to guarantee the positivity are also the sufficient and necessary condition that the quadratic programming problem (\ref{Quadratic programming}) has an interior solution.

When  the assumption $1+\sum_{j=1}^{k}(\theta_j-\theta_i)(\theta_j-\phi_u)/{\sigma_j^2} > 0$ does not hold for some $i=1, \dots, k$, there is no explicit solution  for the optimal weights.
In such settings, the computational methods such as interior point and conjugate gradient can be used to find the numerical solutions.

\vskip 3pt
\noindent

\end{document}